\documentclass[twocolumn,showpacs,preprintnumbers,amsmath,amssymb]{revtex4}

\newtheorem{theorem}{Theorem}

\newtheorem{lemma}[theorem]{Lemma}
\newtheorem{corollary}[theorem]{Corollary}

\newtheorem{Prop}[theorem]{Proposition}

\usepackage{graphicx,dcolumn,bm,color,comment}
\usepackage{hyperref}

\newcommand{\tr}{\mathop{\mathrm{tr}}\nolimits} 
\newcommand{\E}{\mathop{\mathcal{E}}\nolimits}
\newcommand{\RA}{\mathop{\mathcal{R}}\nolimits}
   \newcommand{\HA}{\mathop{\mathcal{H}}\nolimits} 
 \newcommand{\LA}{\mathop{\mathcal{L}}\nolimits}   
    \newcommand{\CA}{\mathop{\mathbb{C}}\nolimits} 
\newcommand{\I}{\mathop{\mathbb{I}}\nolimits}   
\newcommand{\IA}{\mathop{\mathcal{I}}\nolimits}
\newcommand{\bra}[1]{\langle #1 |}
\newcommand{\ket}[1]{| #1 \rangle}
\newcommand{\bracket}[2]{\langle #1 | #2 \rangle}
\newcommand{\ketbra}[2]{| #1 \rangle \langle #2 | }

\begin{document}
\sloppy
\title{
Solution to the mean king's problem using quantum error-correcting codes
}

\author{Masakazu YOSHIDA$\ ^{1}$}
\email{masyoshi@mail.doshisha.ac.jp}
\author{Gen KIMURA$\ ^{2}$}
\email{gen@shibaura-it.ac.jp}
\author{Takayuki MIYADERA$\ ^{3}$}
\author{Hideki IMAI}
\author{Jun Cheng$\ ^{1}$}

\affiliation{
$\ ^{1}$
Dept. of Intelligent Information Eng. and Sci., Doshisha University. 
\\
1-3 Tatara Miyakodani, Kyotanabe-shi, Kyoto, 610-0394, Japan. 
}

\affiliation{
$\ ^{2}$
College of Systems Engineering and Science, Shibaura Institute of Technology, 
\\307 Fukasaku, Minuma-ku, Saitama-shi, Saitama, 337-8570, Japan.
}

\affiliation{
$\ ^{3}$
Department of Nuclear Engineering, Kyoto University,
\\
Kyoto, 615-8540, Japan. 
}

\date{\today}

\begin{abstract}
We discuss the so-called mean king's problem, a retrodiction problem among non-commutative observables, in the context of error detection. 
Describing the king's measurement effectively by a single error operation, we give a solution of the mean king's problem using quantum error-correcting codes. 
The existence of a quantum error-correcting code from a solution is also presented. 
\end{abstract}

\pacs{03.67.Hk, 03.67.Pp, 03.65.Ud, 03.65.Aa}

\maketitle

\section{Introduction}
In 1987, Vaidman, Aharonov, and Albert formulated the mean king's problem \cite{VAA87} to provide a way to discriminate eigenstates among noncommutative observables with the help of classical delayed information. 
The problem is often told as a tale \cite{AE01,EA01,KR05} that a mean king gives a physicist, say Alice, the challenge for that discrimination problem. 
In the tale, the king performs a projective measurement of one of the observables $\sigma_x,\sigma_y,$ and $\sigma_z$ on a qubit system prepared by Alice.   
Alice is given an opportunity to measure the system, and then the king reveals the observable he has measured. 
Immediately after that, Alice is required to correctly guess the king's outcome. 
We say that the king's problem has a solution if Alice can find a strategy to be successful in this challenge.  
In \cite{VAA87}, inspired by the Aharonov-Bergmann-Lebowitz rule \cite{ABL64}, a solution is constructed by making use of an entanglement between the measured qubit and another qubit secretly kept by Alice.  

The problem has been generalized in several directions: 
Most naturally, a quantum system with $d \ge 2$ levels has been considered with the king's measurement being one of the complete (i.e., $d+1$ numbers of) mutually unbiased bases (MUBs) \cite{WF88,Iv81}. 
With particular constructions of MUBs, solutions have been successfully shown for $d=3$ \cite{AE01}, $d=$ prime \cite{EA01}, and $d=$ power of prime \cite{Ara2003}. 
For a general $d$, the existence of the solution is shown to be that of the orthogonal Latin squares, irrespective of the way of the construction of MUBs but under the restriction of Alice's measurement to be 
a projection valued measure (PVM) measurement \cite{HHH205}. 
This implies that for some cases, e.g., $d=6$, we have no solutions to the problem.
However, allowing Alice to perform a positive operator valued measure (POVM) measurement, it has been shown that a solution always exists for arbitrary dimension \cite{KTO06}. 
On the other hand, a non-MUB measurement for the king's measurements has been considered for $d=2$ \cite{M89,HHH05} and in a general dimension in \cite{RW07}. 
As a different line of generalization, it is shown that there are no solutions 
if Alice does not use the blessings of entanglement \cite{KTO07,Ar03}. 
Recently, by showing the relationship between the mean king's problem and finite dual affine plain geometry, a solution to the problem is introduced in the odd prime dimensional case \cite{Re125406}. 
Connected to the above relationship, it is considered that Alice guesses a measurement employed by the king 
without any classical delayed information revealed by the king. 
This derivation is called tracking the king \cite{Re121704}. 

In this paper, we investigate the king's problem from the view point of error detection and correction. 
By considering the king's measurement as an error, the problem becomes a certain kind of an error detection problem.  
To see the idea, let us see how the solution is constructed in \cite{VAA87}:  
In Alice's preparation, she utilizes an entangled state on two qubits, one of which is for the king and the other is kept secretly on Alice's hand. 
In the context of error correction, this corresponds to finding a good coding system, especially by adding a redundant system to have the error resilience. 
Then, Alice finds an observable, the measurement of which makes her successful in guessing the king's outcome. 
The key fact here is that the observable has the orthogonal eigenspaces such that the post-measurement state for each outcome of any Pauli matrix on the king's space belongs to one of the eigenspaces.  
The general idea to construct a reliable error detection method is to find an orthogonal decomposition on the larger Hilbert space so that each error-state belongs to a different subspace. 
Thus, one can consider Alice's measurement as the syndrome measurement to diagnose the error, corresponding to the king's outcome, in the context of the mean king's problem. 
In this way, the king's problem is fairly translated to that of the error detection. 

Notice, however, the existence of the delayed information of the king's measurement basis makes the problem complicated and one cannot apply the general theory of error correction straightforwardly. 
Nevertheless, we show that the king's problem can be described by a single error operation described by the collection of error operators $(L_k)_k$ which effectively describes the king's measurements and realizes the above mentioned error correcting strategy. 
For the setting in \cite{VAA87}, we find a collection of error operators $(L_k)_k$ such that each of the operators projects a Bell state to orthogonal subspace and a particular combination of them corresponds to an eigen-projections of each Pauli matrix. 
Then, Alice can perform a syndrome measurement corresponding to the orthogonal subspace and can correctly guess the king's outcome according to the combination rule connecting the error operators and the choice of the king's measurement.  

This paper is organized as follows. 
In Sec. II, we review quantum error-correcting codes and the general condition for the existence of the codes. 
In Sec. III, we introduce the mean king's problem in the most general setting using measurement operators 
and give a solution to the problem which consists of analogical error detection and error correction. 
In Sec. IV, the existence of quantum error-correcting codes is discussed in the case where there exist solutions to the problem. 
In Sec. V, problems solvable with our method are constructed from any orthonormal basis. 
In Sec. VI, we discuss higher dimensional codes as the solutions to the problem. 
Finally, we summarize this paper in Sec. VII. 

\section{Review of Quantum Error-Correcting Codes}

Throughout this paper, we treat a finite dimensional Hilbert space and we regard Hilbert spaces (resp. density operators) 
in the same light as quantum systems (resp. quantum states). 
Let $\rho$ be a density operator on a Hilbert space $\HA$, i.e., $\rho\geq 0$ and $\mathrm{tr}\rho =1$. 
We denote by ${\cal S}({\cal H})$ the set of density operators on $\HA$. 
A general quantum operation is described by a trace non-increasing completely positive (CP) map $\Lambda$, which is represented by a collection of linear operators $(A_i)_i$ such that $\Lambda(\rho)=\sum_iA_i\rho A_i^{\dagger}$. 
A trace non-increasing condition reads $\sum_i A_i^\dagger A_i \le \I$ while the equality hold iff it is trace preserving. 
This representation is called a Kraus representation and each $A_i$ is called a Kraus operator. 
In the context of noise operation, $A_i$ is called a noise operator. 
In the following, we often use an abbreviation $\Lambda = (A_i)_i$ to imply the Kraus representation $\Lambda(\rho) = \sum_i A_i \rho A^\dagger_i$.    

Let $\E = (E_i)_i$ be an error operation on a $d$ dimensional Hilbert space. 
An $n$ dimensional subspace $C \subset \HA$ is called a $(d,n)$ quantum error-correcting code against $\E = (E_i)_i$ if there exists a trace preserving CP map $\RA = (R_j)_j$ such that 
\begin{equation*}
\RA \circ \E (\rho) \propto \rho
\end{equation*}
for any state $\rho$ whose support lies in $C$. 
The general condition for the existence of the error-correcting code was given by Knill-Laflamme \cite{KL00}.
\begin{theorem}\label{KL} A necessary and sufficient condition for $C \subset \HA$ to be a $(d,n)$ quantum code against an error operation $\E = (E_i)_i$ is that 
$$
P E_i^{\dagger} E_{i'} P =\lambda_{ii'}P \hspace{0.3cm} \forall i,i'
$$
with a positive matrix $(\lambda_{ii'})_{i,i'}$ where $P$ denotes the projector onto $C$. 
\end{theorem}
Moreover, it is easy to see that the error correcting code $C$ can also correct an error operation spanned by $\{ E_i\}_i$ \cite{NC}. 

Notice that in general the orthogonality of the code states with error is only sufficient but not necessary. 
However, in the following, we use the orthogonality condition to connect the mean king's problem and error detecting problem. 
 
\section{Solution to the mean king's problem using quantum error-correcting codes}

The essence of the mean king's problem is summarized as follows: 

(i.) The king performs one of the measurements on a quantum state $\HA_K$, where the initial state is prepared by Alice.  

(ii.) Alice is then allowed to perform a measurement on the system. 

(iii.) Immediately after the king reveals the measurement type he performed, Alice is required to answer the king's measurement outcome. 

Given the set of the king's measurements, a solution to the mean king's problem is defined as a pair of an initial state $\rho$ and Alice's measurement ${\bf A}$ with which Alice can successfully guess the king's outcome with delayed information $J$ of the king's measurement.
 
Here, we shall give several equivalent conditions of a solution $(\rho,{\bf A})$ to the mean king's problem. 
Let ${\bf J}$, ${\bf I}$, and ${\bf A}$ be random variables (including measurements) for the king's measurement type $J$, the king's outcome $i$, and Alice's outcome $a$. 
We denote the joint probability distribution of ${\bf J},{\bf I},{\bf A}$ by $\mathrm{Pr}(J,i,a)$, and the conditional probability of ${\bf I}$ given ${\bf J} = J, {\bf A} = a$ by $\mathrm{Pr}(i|J,a)$, and so on. Note that we shall omit the dependence of $\rho$ of these quantities for the notational simplicity.  
\begin{Prop}\label{prop:solMKP} The followings are equivalent: Given the king's measurement set, 

{\rm (s1.)} $(\rho,{\bf A})$ is a solution to the mean king's problem. 

{\rm (s2.)} There exists an estimation function $s(J,a)$ such that $\mathrm{Pr}(s(J,a) | J,a) = 1$. 
	
{\rm (s3.)} $H({\bf I} | {\bf J},{\bf A}) = 0$ where $H({\bf I} | {\bf J},{\bf A})$ is the conditional entropy of ${\bf I}$ given ${\bf J}$ and ${\bf A}$. 

{\rm (s4.)} An index set $X^{(J,i)} \supseteq \{ a | \mathrm{Pr}(J,i,a) \neq 0 \}$ satisfies  
\begin{equation}\label{cond:sol}
X^{(J,i)} \cap X^{(J,i')} = \emptyset \hspace{0.3cm} \forall J, \forall i\neq i'. 
\end{equation} 
\end{Prop}

The proof is straightforward (Note that if Alice gets outcome $a$ and heard the king's measurement $J$, then the king's outcome $i$ should satisfy $a \in X^{J,i}$. However, by condition \eqref{cond:sol}, such $i$ uniquely exists, which determines Alice's estimation function.) 

In a quantum setting of the problem, Alice can utilize entanglement between the king's system $\HA_K$ and another quantum system $\HA_A$ kept secretly by Alice. 
Let $d$ and $d'$ be dimensions of $\HA_K$ and $\HA_A$, respectively.  
Although the king's measurement is conventionally restricted to a class of projective measurements, we allow the king to perform a general class of measurement process described by a collection of measurement operators. 
Namely, letting ${\bf M}^{(J)} = (M^{(J)}_i)_i \ (J = 1,\ldots,m)$ be a collection of measurement operators on $\HA_K$ for a $J$th measurement, the post-measurement state from the density operator $\rho$ is given by 
$$
\rho^{(J)}_i = \frac{M^{(J)}_i \rho {M^{(J)}_i}^\dagger }{p^{(J)}_i},
$$
where $i$ is the king's outcome and $p^{(J)}_i := \tr M^{(J)}_i \rho {M^{(J)}_i}^\dagger $ is the probability to get the $i$th outcome. 
Moreover, the most general class of Alice's measurement can be described by POVM measurement ${\bf A} = (A_a)_a$ on $\HA_A \otimes \HA_K$. 

Motivated by Theorem \ref{KL}, we can relate a solution to the general mean king's problem to the error detection problem as follows:   

\begin{theorem}\label{prop:Main}
Let $C \subset \HA_A \otimes \HA_K$ be an $n$ dimensional subspace and $P$ the projection operator onto $C$. 
If there exists a quantum operation $(L_a)_{a}$ on $\HA_K$ and non-empty index sets 
$X^{(J,i)} \subset \{1,2,\ldots,l\}$ satisfying 
\begin{eqnarray*}
&\rm{(c1.)}& {\I_A \otimes}M^{(J)}_i = \sum_{a \in X^{(J,i)}} f^{(J,i)}_{a}{\I_A \otimes} L_a \label{eq:ML} \ { {\rm on} \ C}, \label{eq:Mdec}\\
&\rm{(c2.)}& X^{(J,i)} \cap X^{(J,i')} = \emptyset  \hspace{0.3cm} \forall J, \forall i \neq i', \label{eq:I} \\ 
&\rm{(c3.)}&P(\I_A \otimes L_a)^\dagger (\I_A \otimes L_{a'}) P = \lambda_{a} \delta_{a a'}P, \label{eq:OC}
\end{eqnarray*}
for some $\lambda_{a} \ge 0$ and $f^{(J,i)}_{a} \in \CA$, then 

\rm{(i.)} by using any code state in $C$ as an initial state, 
Alice can guess the king's outcome perfectly. 

\rm{(ii.)} $C$ is a quantum error-correcting code 
against $\mathrm{span}\{\I_A \otimes L_a\}_{a}$. 
\end{theorem}

The theorem tells us that we can effectively describe the king's measurements introducing a single error operation $(L_a)_a$ through the decompositions (c1). 
Note that condition (c3) is a sufficient condition for 
distinguishing the kinds of error operators perfectly on $C$ 
and implies that $C$ is a quantum error-correcting code against the error operation. 

{\bf Proof.} \
(i.) Let $|\Phi\rangle\in C$ be an initial state prepared by Alice. 
If the king chooses the $J$th measurement 
and obtains the $i$th outcome, 
then condition (c1) implies that the post measurement state is proportional to 
$$
|{\mathbb I}_A\otimes M_i^{(J)}\Phi\rangle\in\bigoplus_{a \in X^{(J,i)}}{\cal A}_a,
$$
where ${\cal A}_a$ is a subspace spanned by $\{ {\mathbb I}_A\otimes L_a C\}(a=1,2,\ldots ,l)$; note here that,  
from condition (c3), ${\cal A}_a$ and ${\cal A}_{a'}$ are orthogonal subspaces for any $a\neq a'$. 
Let $P_a$ be the projection operator onto ${\cal A}_a$, 
then ${\cal P}:=(P_1,P_2,\ldots ,P_l,P^{\perp})$ 
forms a discrete PVM, 
where $P^{\perp}:={\mathbb I}_A\otimes{\mathbb I}_K-\sum_{a=1}^{l}P_a$. 

Let Alice performs the PVM measurement ${\cal P}$ and obtains outcome $a$. 
With the revealed $J$ by the king, Alice is assured that the king's outcome $i$ satisfied $a \in X^{(J,i)}$. 
Thus Alice can correctly guess the king's outcome [see (s4) of Proposition \ref{prop:solMKP}]. 

(ii) Condition (c3) implies that $C$ is a quantum error-correcting code 
against the error operation $\{ {\mathbb I}_A\otimes L_a\}_{a =1}^{l}$ from Theorem \ref{KL}. 
In addition, 
$C$ can also correct an error operation $\mathrm{span}\{ {\mathbb I}_A\otimes L_a\}_{a =1}^{l}$. 
\hfill $\blacksquare$

\section{Existence of quantum error-correcting codes}
In this section, we discuss conversely the existence of quantum error-correcting codes 
against error operators which consist of operators derived from the measurements 
if there exists a solution to the mean king's problem. 

Let $\HA_A=\HA_K := \HA$ and an entangled quantum state (in the form of the Schmidt decomposition): 
\begin{equation}\label{eq:Sch}
\ket{\Psi_{\bm \eta}} = \sum_{j=1}^d \eta_j \ket{\psi_j} \otimes \ket{\phi_j} \ (\eta_j > 0), 
\end{equation}
be prepared as an initial state with orthonormal bases $\{\ket{\psi_j}\}_j$ and $\{\ket{\phi_j}\}_j$ of $\HA$ satisfying $\sum_j\eta_j^2=1$. 
A maximal entangled state is $\ket{\Psi_{\bm \eta}} $ with ${\bm \eta} = (\eta_1,\ldots,\eta_d ) = (\frac{1}{\sqrt{d}},\ldots,\frac{1}{\sqrt{d}})$. 
In the following, we shall denote a maximal entangled state by $\ket{\Psi} $. 

Let $\LA(\HA)$ be the set of linear operators on $\HA$. 
Noting that $\eta_j > 0$ for any $j$, it is easy to see that $\LA(\HA)$ is a $d^2$ dimensional Hilbert space by introducing the following inner product dependent on $\ket{\Psi_{\bm \eta}}$: 
\begin{eqnarray}
\bracket{A}{B}_{Sc} &:=& d \bracket{\I\otimes A\Psi_{\bm \eta}}{\I\otimes B\Psi_{\bm \eta}} \label{eq:IP}\\ 
&=& d \sum_j \eta_j^2 \bra{\phi_j}A^\dagger B \ket{\phi_j}. \nonumber
\end{eqnarray}
One can introduce the following isomorphism between $\LA ({\HA})$ and $\HA\otimes\HA$ (which is similar to the Choi-Jamio\l kowski isomorphism): 
\begin{equation}\label{eq:iso}
L\in\LA ({\HA})\mapsto {\cal I}_{{\bm \eta}}(L) := \sqrt{d} \ket{\I\otimes L\Psi_{\bm \eta}}\in\HA\otimes\HA   
\end{equation}
in a way to preserve the inner products of $\LA ({\HA})$ and $\HA\otimes\HA$. 
Indeed, we have $\bracket{\IA_{\bm \eta} (L)}{\IA_{\bm \eta} (L')}
=d \sum_{i,j}\eta_i \eta_j \bracket{\psi_i}{\psi_j}\bracket{\phi_i}{L^{\dag}L'\phi_j}
=d \sum_{j} \eta_j^2 \bracket{\phi_j}{L^{\dag}L'\phi_j} =\bracket{L}{L'}_{Sc}$. 
Note that Eq. \eqref{eq:IP} is a generalization of the Hilbert-Schmidt inner product: 
$$
\bracket{A}{B}_{HS}:= \tr A^\dagger B
$$ 
which is the case when $\ket{\Psi_{\bm \eta}}$ is a maximal entangled state $\ket{\Psi}$. 
We denote the isomorphism \eqref{eq:iso} for the maximal entangled state just by ${\cal I}$. 

\begin{lemma}\label{lem:comp}
For any (unnormalized) orthogonal basis $\{L_a\}_{a=1}^{d^2}$ of $\LA(\HA)$ such that $\bracket{L_a}{L_{a'}}_{Sc} = \alpha \delta_{aa'} \ (\alpha > 0)$,
we have 
\begin{equation}\label{eq:comp}
\sum_{a=1}^{d^2} L_a^\dagger L_a = \sum_{j=1}^d\frac{\alpha}{\eta_j^2}\ketbra{\phi_j}{\phi_j}.  
\end{equation}
If $\ket{\Psi_{\bm \eta}}$ is a maximal entangled state, we have 
$$
\sum_{a=1}^{d^2} L_a^\dagger L_a = \alpha d \I. 
$$
\end{lemma}

{\bf Proof.} \ 
To prove Eq.~\eqref{eq:comp}, it is enough to observe that $\sum_{a=1}^{d^2} L_a^\dagger L_a$ is the inverse of $\sum_{j=1}^d\frac{\eta_j^2}{\alpha}\ketbra{\phi_j}{\phi_j}$. 
Note that $\ketbra{\chi}{\xi} \ (\forall \ket{\xi},\ket{\chi} \in \HA)$ can be written as $\ketbra{\chi}{\xi}=\sum_a 1/\alpha\bracket{L_a}{\ketbra{\chi}{\xi}}_{Sc} L_a
=\sum_{a,j}d\eta_j^2/\alpha \bracket{\xi}{\phi_j} \bra{\phi_j}L_a^{\dagger}\ket{\chi} L_a$ by the orthogonal basis $\{L_a\}$.   
Applying an orthonormal basis $\{\ket{\chi_k}\}_k$ of $\HA$ to $\ket{\chi}$ of this equality, 
we get $\bra{\xi} = \bra{\xi} (\sum_{j} \frac{\eta_j^2}{\alpha} \ketbra{\phi_j}{\phi_j} (\sum_a L_a^\dagger L_a)$ 
by taking inner product $\bra{\chi_a}$ and summing them over $a$. 
Since $\bra{\xi}$ is an arbitrary vector, 
we have $(\sum_{j} \frac{\eta_j^2}{\alpha} \ketbra{\phi_j}{\phi_j}) (\sum_a L_a^\dagger L_a) = \I$. 
\hfill $\blacksquare$

Using Lemma \ref{lem:comp}, we obtain the following theorem which implies the existence of a quantum error-correcting code $C$ 
against an error operation $(L_a)_a$ from a solution of the mean king's problem. 

\begin{theorem}\label{prop:Mains}
Let ${\cal P}=(P_a=\ketbra{p_a}{p_a})_{a=1}^{d^2}$ be a PVM on $\HA_A\otimes\HA_K$ with an orthonormal basis $\{ \ket{p_a}\}_{a=1}^{d^2}$. 
If the initial state $\ket{\Psi_{\bm \eta}}$ in \eqref{eq:Sch} with the PVM measurement $\hat{\cal P}$ provides a solution to the mean king's problem, 
there exists a quantum operation $(L_a)_{a=1}^{d^2}$ on $\HA_K$ and index sets $X^{(J,i)}$ satisfying the following conditions, 
\begin{eqnarray*}
&\rm{(c1')}& M^{(J)}_i = \sum_{a \in X^{(J,i)}} f^{(J,i)}_{a} L_a, \\
&\rm{(c2')}& X^{(J,i)} \cap X^{(J,i')} = \emptyset \hspace{0.3cm} \forall J, \forall i \neq i', \\
&\rm{(c3')}& \bracket{L_a}{L_{a'}}_{Sc} = \alpha \delta_{aa'},  
\end{eqnarray*}
for some $\alpha > 0$ and $f^{(J,i)}_{a} \in \CA$. 
For the case of a maximal entangled state $\ket{\Psi}$, $(L_a)_a$ can be chosen to satisfy the trace preserving condition, i.e., $\sum_a L_a^\dagger L_a = \I_K$, with $\alpha = 1/d$. 
\end{theorem}

Note that conditions (c1')-(c3') are parallel to (c1)-(c3) in Theorem \ref{prop:Main}: (c1') is an equality condition not only on the code space $C$; 
(c3') implies the orthogonality condition (c3) on $C$. 
Therefore, Theorem \ref{prop:Mains} gives a ``reverse" statement of Theorem \ref{prop:Main}.

{\bf Proof.} \ First, using isomorphism \eqref{eq:iso}, we define an error operation $( L_a)_{a=1}^{d^2}$ on $\HA_K$ 
by 
$$
L_a := {\cal I}_{\bm \eta}^{-1} (\sqrt{\alpha} \ket{p_a}) \ (\Leftrightarrow \ket{p_a} = \sqrt{\frac{d}{\alpha}}\ket{\I \otimes L_a \Psi_{\bm \eta}}), 
$$ 
for some $\alpha >0$. 
Then, orthogonality condition (c3') follows from an inner preserving property of the isomorphism. 
Moreover, Lemma \ref{lem:comp} implies $\sum_{a}L_a^{\dagger}L_a\leq\I_K$ if we put $\alpha :=\min_{j}\{\eta_j^2\}_j > 0$. 
For the case of the maximal entangled state, $\alpha = 1/d$ and we have $\sum_a L_a^\dagger L_a = \I_K$.  

Next, the conditional probability corresponding to the random variables 
has the following form: 
\begin{equation*}
P(J,i,a)=P(J)\bracket{\I_A\otimes M_{i}^{(J)}\Psi_{\bm \eta}}{P_a(\I_A\otimes M_{i}^{(J)})\Psi_{\bm \eta}}, 
\end{equation*}
where 
\begin{eqnarray*}
&& \bracket{\I_A\otimes M_{i}^{(J)}\Psi_{\bm \eta}}{ P_a(\I_A\otimes M_{i}^{(J)})\Psi_{\bm \eta}} \nonumber\\
&=& \frac{d}{\alpha}|\bracket{\I_A\otimes M_{i}^{(J)}\Psi_{\bm \eta}}{\I_A\otimes L_a\Psi_{\bm \eta}}|^2 = \frac{1}{d}\alpha|\bracket{M_{i}^{(J)}}{L_a}_{Sc}|^2. \nonumber\label{coef}
\end{eqnarray*}
Since $\ket{\Psi_{\bm \eta}}$ and ${\cal P} = (P_a)$ is the solution to the problem, an index set $X^{(J,i)}:=\{ a\mid \bracket{M_{i}^{(J)}}{L_a}_{Sc} \neq 0\} \supseteq \{ a | \mathrm{Pr}(J,i,a) \neq 0 \}$ satisfies condition (c2') [see (s4) of Proposition \ref{prop:solMKP})]. 
Therefore,  
\begin{eqnarray*}
M_{i}^{(J)} &=& \sum_a\frac{1}{\alpha}\bracket{L_a}{M_{i}^{(J)}}_{Sc}L_a\\
&=& \sum_{a\in X^{(J,i)}}\frac{1}{\alpha}\bracket{L_a}{M_{i}^{(J)}}_{Sc}L_a 
\end{eqnarray*}
holds. This proves condition (c1'). 
\hfill $\blacksquare$

\section{Class of Problems solved with quantum error-correcting codes}\label{Const} 
We construct a pair of a set of operators and index sets 
which satisfies the conditions (c1)-(c3) for a maximal entangled state from any orthogonal basis. 
We also show settings of the mean king's problem with measurements which consist of the pair. 
The settings are solved with the method of Theorem \ref{prop:Main} by using quantum error-correcting codes. 

Let $\HA_K=\HA_A:=\HA$ be a $d$ dimensional Hilbert space and 
$\{\ket{f_i}\}_{i=1}^d$ an orthonormal basis for the space. 
We define operators $( L_{ij})_{i,j=1}^d$ on $\HA_K$ by 
\begin{eqnarray*}
L_{ij}:=\IA^{-1}(\frac{1}{\sqrt d} \ket{f_i}\otimes\ket{f_j}), \label{Isom}
\end{eqnarray*}
where we use the isomorphism $\IA$ for a maximal entangled state 
$1/\sqrt d\sum_{i=1}^d\ket{f_i}\otimes\ket{f_i}$. 
$(L_{ij})_{i,j}$ is an orthogonal basis with $\bracket{L_{ij}}{L_{i'j'}}_{HS}=\frac{1}{d}\delta_{(i,j)(i',j')}$ (this implies (c3)), 
then $\sum_{i,j}L_{ij}^\dagger L_{ij} = \I$ holds from Lemma \ref{lem:comp}. 

We define index sets 
$X^{(J,i)}:=\{ (l,J^{(i)}(l))\}_{l\in [d]}\subset [d]\times [d]$ ($J=1,2,\dots m$ and $i\in [d]$) 
by $\{ J^{(i)}(l)\}_{l\in [d]} = [d]$ and $J^{(i)}(l)\neq J^{(i')}(l)$ for any $i\neq i'$ and $l$, where $[d]:=\{1,2,\dots , d\}$. 
For $J=0$, we define $X^{(0,i)}:=\{ (i,l)\}_{l\in [d]}\subset [d]\times [d]$ ($i\in [d]$). 
Then, $X^{(J,i)}\cap X^{(J,i')}=\emptyset$ holds for any $J$ and $i\neq i'$ (this implies (c2)).
Remark that a size $d\times d$ matrix 
$\hat J=(J_{il}:=J^{(i)}(l))_{1\leq i,l\leq d} (J=1,2,\ldots ,m)$ 
is a Latin square, 
i.e., $\hat J$ has $d$ different symbols, 
each occurring exactly once in each row and each column. 

Now we construct a collection of measurement operators from the operators $(L_{ij})_{i,j}$ and the index sets $X^{(J,i)}$. 
\begin{corollary}
We define $M^{(J)}:=(M_i^{(J)}:=\sum_{(j,k)\in X^{(J,i)}}L_{jk})_{i\in [d]}$ ($J=1,2,\ldots ,m$) 
and $M^{(0)}:=(M_i^{(0)}:=\sum_{j=1}^dL_{ij})_{i\in [d]}$. 
Then, $M^{(J')}$ is a collection of measurement operators 
with $\sum_{i\in [d]}{M_i^{(J')}}^{\dagger}M_i^{(J')}=\I$ 
for any $J'\in\{0,1,\ldots ,m\}$. 
\end{corollary}

That is, 
we can solve the mean king's problem in which the king employs $M^{J}(J=0,1,\ldots ,m)$ 
by using the method in Theorem \ref{prop:Main} 
since $( L_{ij})_{i,j}$ and $X^{(J,i)}$ satisfy the conditions (c1)-(c3) 
for $M^{J}$ and the maximal entangled state. 

{\bf Proof.} \ 
For any $i$ and $J\neq 0$, 
$\ket{M_i^{(J)}f_j}=\frac{1}{\sqrt d}\ket{f_{J^{(i)}(j)}}$ holds because 
$\sum_{j}\ket{f_j}\otimes\ket{M_i^{(J)}f_j} 
= \IA (M_i^{(J)}) 
= \sum_{(j,k)\in X^{(J,i)}}\IA (L_{jk}) 
= \sum_{(j,k)\in X^{(J,i)}}\frac{1}{\sqrt d}\ket{f_j}\otimes\ket{f_k}
= \sum_{j}\ket{f_j}\otimes\frac{1}{\sqrt d}\ket{f_{J^{(i)}(j)}}$. 
Then, we observe 
\begin{eqnarray*}
\bracket{f_l}{\sum_{i=1}^d{M_i^{(J)}}^{\dagger}M_i^{(J)}f_{l'}}
&=&\sum_{i,k=1}^d\overline{\bracket{f_k}{M_i^{(J)}f_l}}\bracket{f_k}{M_i^{(J)}f_{l'}}\\
&=&\frac{1}{d}\sum_{i,k=1}\bracket{f_{J^{(i)}(l)}}{f_k}\bracket{f_k}{f_{J^{(i)}(l')}}\\
&=&\frac{1}{d}\sum_{i=1}\bracket{f_{J^{(i)}(l)}}{f_{J^{(i)}(l')}}
=\delta_{ll'}. 
\end{eqnarray*}

For any $j,k\in[d]$, 
$\ket{M_i^{(0)}f_l}=\frac{\delta_{il}}{\sqrt d}\ket{f}$ (where $\ket{f}:=\sum_j\ket{f_j}$) holds because 
$\sum_l\ket{f_l}\otimes M_i^{(0)}\ket{f_l} 
= \IA (M_i^{(0)}) 
= \sum_j \IA (L_{ij}) 
= \sum_j \frac{1}{\sqrt d}\ket{f_i}\otimes\ket{f_j} 
= \ket{f_i}\otimes\frac{1}{\sqrt d}\sum_j\ket{f_j} 
= \sum_l\ket{f_l}\otimes\frac{\delta_{il}}{\sqrt d}\sum_j\ket{f_j}$. 
Then, we observe 
\begin{eqnarray*}
\bracket{f_l}{\sum_{i=1}^d{M_i^{(0)}}^{\dagger}M_i^{(0)}f_{l'}}
&=&\sum_{i,k=1}^d\frac{1}{d}\delta_{il}\delta_{il'}
\overline{\bracket{f_k}{f}}\bracket{f_k}{f}\\
&=&\frac 1 d\sum_{i=1}^d\delta_{il}\delta_{il'}\bracket{f}{f}
=\delta_{ll'}.
\end{eqnarray*}
This implies that $\sum_{i\in [d]}{M_i^{(J')}}^{\dagger}M_i^{(J')}=\I$ holds for any $J'\in\{ 0,1,\ldots ,m\}$. 
\hfill $\blacksquare$

\section{Toward Higher Dimensional Quantum Error-Correcting Codes}\label{Const2}
So far, we have only seen the solutions with one- dimensional (1D) quantum codes. 
In the following example, 
we introduce a problem 
which are solved by using three-dimensional (3D) quantum code. 

First we construct a problem in which the king employs two kinds of projective measurements in a two-dimensional (2D) Hilbert space in preparation for the example 
and show that the 1D dimensional quantum code spanned by the Bell state $\ket{\Psi^+}$ is a solution to the problem. 
Let ${\cal H}_A$ and ${\cal H}_K$ be 2D dimensional Hilbert spaces. 
We define operators $(\hat L_a)_{a=1}^4$ 
with $\sum_{a=1}^4\hat L_a^{\dagger}\hat L_a=\mathbb I$ by 
$\hat L_1:=X_0Z_0, \hat L_2:=X_1Z_0, 
\hat L_3:=X_0Z_1, \hat L_4:=X_1Z_1$, 
where $X_0:=\ket{+}\bra{+}, X_1:=\ket{-}\bra{-}$ and $Z_0:=\ket{0}\bra{0}, Z_1:=\ket{1}\bra{1}$ are eigen-projectors of $\sigma_x$ and $\sigma_z$, respectively. 
For projective measurements $\hat M^{(1)}:=(X_0,X_1)$, $\hat M^{(2)}:=(Z_0,Z_1)$, 
and $\ket{\Psi^+}$, we observe 
\begin{eqnarray*} 
&&X_0=\hat L_1+\hat L_3,\hspace{0.3cm}X_1=\hat L_2+\hat L_4,\\
&&Z_0=\hat L_1+\hat L_2,\hspace{0.3cm}Z_1=\hat L_3+\hat L_4,
\end{eqnarray*}
and 
$$\bracket{\IA(\hat L_{a})}{\IA(\hat L_{a'})} = \bracket{\hat L_{a}}{\hat L_{a'}}_{HS}=\frac{\delta_{aa'}}{2}.$$ 
Therefore, the operators $(\hat L_a)_{a=1}^4$ and 
the index sets from the relation between $(\hat L_a)_{a=1}^4$ and $(\hat M^{(1)},\hat M^{(2)})$ 
satisfy the conditions (c1)-(c3) for the measurements and a $(4,1)$ quantum code spanned by $\ket{\Psi^+}$. 

In the following, we show a problem solved with a $3$ dimensional quantum code. 
Let $\{\ket{i}\}_{i=0}^2$ be the computational basis of $\CA^3$. 
We define 
$\tilde L_1:=\tilde{X_0}\tilde{Z_0},\tilde L_2:=\tilde{X_1}\tilde{Z_0},
\tilde L_3:=\tilde{X_0}\tilde{Z_1},\tilde L_4:=\tilde{X_1}\tilde{Z_1},
\tilde L_5:=\ket 2\bra 2$, 
where 
\begin{eqnarray*}
&&
\tilde{X_0}:=
\left( \hspace{-\arraycolsep}
\begin{array}{ccc}
 &  & 0 \\
\multicolumn{2}{c}{\raisebox{1.0ex}[0pt]{\Large{${X_0}$}}} & 0  \\
0 & 0 & 0
\end{array}
\right),\hspace{0,3cm}
\tilde{X_1}:=
\left( \hspace{-\arraycolsep}
\begin{array}{ccc}
 &  & 0 \\
\multicolumn{2}{c}{\raisebox{1.0ex}[0pt]{\Large{${X_1}$}}} & 0  \\
0 & 0 & 0
\end{array}
\right),\\
&&
\tilde{Z_0}:=
\left( \hspace{-\arraycolsep}
\begin{array}{ccc}
 &  & 0 \\
\multicolumn{2}{c}{\raisebox{1.0ex}[0pt]{\Large{${Z_0}$}}} & 0  \\
0 & 0 & 0
\end{array}
\right),\hspace{0,3cm}
\tilde{Z_1}:=
\left( \hspace{-\arraycolsep}
\begin{array}{ccc}
 &  & 0 \\
\multicolumn{2}{c}{\raisebox{1.0ex}[0pt]{\Large{${Z_1}$}}} & 0  \\
0 & 0 & 0
\end{array}
\right).
\end{eqnarray*}
Then, $\sum_a\tilde L_a^{\dagger}\tilde L_a=\mathbb I$ holds. 
Let $\tilde C$ be a $(9,3)$ quantum code spanned by 
$\{\ket i\otimes (\ket 0+\ket 2)\mid i\in\{0,1,2\}\}$, 
then, 
$$\bra i(\bra 0+\bra 2)(\mathbb I\otimes \tilde L_a)^{\dagger}(\mathbb I\otimes \tilde L_{a'})\ket j(\ket 0+\ket 2)=\lambda_{a}\delta_{aa'}\delta_{ij}$$
holds, where $\lambda_{1}=\lambda_{2}=\lambda_{5}=1, \lambda_{3}=\lambda_{4}=0$. 
Therefore, there exists a solution with any code state in $\tilde C$ to the mean king's problem which consists of a collection of measurement operators constructed from $(\tilde L_a)_{a=1}^5$ and suitable index sets. 
Here we present an example of such measurement operators: 
$$\tilde M^{(J)}:=(\tilde M^{(J)}_1,\tilde M^{(J)}_2) \hspace{0.3cm} (J=1,2,3,4),$$ 
where
\begin{eqnarray*}
&&\tilde M^{(1)}_1:=\tilde L_1+\tilde L_2=\tilde Z_0,\\
&&\tilde M^{(1)}_2:=\tilde L_3+\tilde L_4+\tilde L_5=\tilde Z_1+\ket 2\bra 2,\\
&&\tilde M^{(2)}_1:=\tilde L_1+\tilde L_2+\tilde L_5=\tilde Z_0+\ket 2\bra 2,\\
&&\tilde M^{(2)}_2:=\tilde L_3+\tilde L_4=\tilde Z_1,\\
&&\tilde M^{(3)}_1:=\tilde L_1+\tilde L_3=\tilde X_0,\\
&&\tilde M^{(3)}_2:=\tilde L_2+\tilde L_4+\tilde L_5=\tilde X_1+\ket 2\bra 2,\\
&&\tilde M^{(4)}_1:=\tilde L_1+\tilde L_3+\tilde L_5=\tilde X_0+\ket 2\bra 2,\\
&&\tilde M^{(4)}_2:=\tilde L_2+\tilde L_4=\tilde X_1. 
\end{eqnarray*}

\section{Conclusion}\label{Concl}
In this paper, we showed the sufficient condition for solving the mean king's problem 
which consists of the measurements operators by using quantum error-correcting codes. 
In the context of quantum error-correcting codes, 
the orthogonality of the error operators is helpful for error detection and error correction. 
We apply the orthogonality to obtain auxiliary information about the king's outcome 
and the outcome is guessed perfectly with the delayed information. 
It is  shown that there exists such a quantum error-correcting code 
if there exist solutions with a bipartite system to the problem and the one-rank PVM. 
Furthermore, we show the settings of the mean king's problem 
which are solved by using our method with quantum error-correcting codes 
and discuss the possibility of the construction of higher dimensional codes. 

\appendix
\section{Solution to the conventional case in the qubits setting}
In the conventional mean king's problem \cite{VAA87}, 
we show a set of operators, index sets, and a quantum code which satisfy (c1)-(c3). 
Suppose that Alice prepares qubit systems 
$\HA_A\otimes\HA_K\simeq{\mathbb C}^2\otimes{\mathbb C}^2$
in a Bell state 
$\ket{\Psi^+}:=1/\sqrt 2(|0\rangle \otimes|0\rangle +|1\rangle \otimes|1\rangle )$, 
where $|0\rangle :=(1,0)^T$ and $|1\rangle :=(0,1)^T$. 
The king chooses one of the measurements constructed from Pauli matrices $\sigma_x, \sigma_y$, and $\sigma_z$:
\begin{eqnarray*}
&M^{(1)}&:=(M_1^{(1)}:=|+\rangle\langle +|,M_2^{(1)}:=|-\rangle\langle -|),\\
&M^{(2)}&:=(M_1^{(2)}:=|+'\rangle\langle +'|,M_2^{(2)}:=|-'\rangle\langle -'|),\\
&M^{(3)}&:=(M_1^{(3)}:=|0\rangle\langle 0|,M_2^{(3)}:=|1\rangle\langle 1|),\
\end{eqnarray*}
where 
$|+\rangle :=1/\sqrt{2} (1,1)^T,|-\rangle :=1/\sqrt{2} (1,-1)^T,|+'\rangle :=1/\sqrt{2} (1,i)^T,$ 
and $|-'\rangle :=1/\sqrt{2} (1,-i)^T$. 
We define 
\begin{eqnarray*}
&&L_1:=
\IA^{-1}(\ket{\Phi_1})=
\frac{1}{4}\left(
\begin{array}{@{\hskip2pt}
cc
@{\hskip2pt}
}
2 & 1-i\\
1+i & 0\\
\end{array}
\right),\\
&&L_2:=
\IA^{-1}(\ket{\Phi_2})=
\frac{1}{4}\left(
\begin{array}{@{\hskip2pt}
cc
@{\hskip2pt}
}
2 & -1+i\\
-1-i & 0\\
\end{array}
\right),\\
&&L_3:=
\IA^{-1}(\ket{\Phi_3})=
\frac{1}{4}\left(
\begin{array}{@{\hskip2pt}
cc
@{\hskip2pt}
}
0 & 1+i\\
1-i & 2\\
\end{array}
\right),\\
&&L_4:=
\IA^{-1}(\ket{\Phi_4})=
\frac{1}{4}\left(
\begin{array}{@{\hskip2pt}
cc
@{\hskip2pt}
}
0 & -1-i\\
-1+i & 2\\
\end{array}
\right),
\end{eqnarray*}
by using the isomorphism $\IA$ for $\ket{\Psi^+}$ 
and a basis measurement $\{\Phi_i\}_{i=1}^4$ employed by Alice in Ref.\cite{VAA87}.  
Then, $\sum_{a=1}^{4}L_{a}^{\dagger}L_{a}={\mathbb I}$ holds 
since $\bracket{L_a}{L_{a'}}_{HS} = \frac 1 2 \delta_{aa'}$ (where this implies (c3)). 
\begin{eqnarray*}
&&M_1^{(1)}=L_1+L_3,\hspace{0.3cm}M_2^{(1)}=L_2+L_4,\\
&&M_1^{(2)}=L_1+L_4,\hspace{0.3cm}M_2^{(2)}=L_2+L_3,\\
&&M_1^{(3)}=L_1+L_2,\hspace{0.3cm}M_2^{(3)}=L_3+L_4,
\end{eqnarray*}
hold. 
That is, the operators $(L_a)_{a=1}^4$ and 
the index sets $X^{(J,i)}$ (see Table \ref{d2})
satisfy conditions (c1) and (c2). 
Therefore, 
we can reconsider the solution \cite{VAA87} to the qubit setting 
from the viewpoint of quantum error-correcting codes. 
\begin{table}[tttt]
\caption{The relationship between the measurement operators and the index sets}\label{d2}
\begin{center}
\begin{tabular}{|c|c|c|}
\hline
$J$ & $i$ & $X^{(J,i)}$ \\ \hline
1 & 1 & $\{1,3\}$ \\
2 & 1 & $\{1,4\}$ \\
3 & 1 & $\{1,2\}$ \\
\hline
\end{tabular}
\begin{tabular}{|c|c|c|}
\hline
$J$ & $i$ & $X^{(J,i)}$ \\ \hline
1 & 2 & $\{2,4\}$ \\
2 & 2 & $\{2,3\}$ \\
3 & 2 & $\{3,4\}$ \\
\hline
\end{tabular}
\end{center}
\end{table}

\section{Construction of measurement operators from computational basis}
We show a collection of measurement operators composed by the computational basis 
as an example of the above construction in Sec. \ref{Const}. 
Let $\{\ket{l}\}_{l=0}^{d-1}$ be the computational basis of $d$ dimensional Hilbert space $\HA_K\simeq {\mathbb C}^{d}$, 
i.e., $l+1$th element (from the top) of $\ket l$ is equal to 1 and the others are equal to $0$. 
We define a $d^2$-tuple of operators 
$(L_{ij})_{i,j=1}^{d}$ by $\IA(\sqrt d L_{ij})=\ket{i-1}\otimes\ket{j-1}$, 
where $1/\sqrt d\sum_{l=0}^{d-1}\ket l\otimes\ket l$ 
is used for the isomorphism. 
We remark that 
each operator $L_{ij}$ is a $d$ by $d$ matrix $(L_{mn}^{(ij)})_{1\leq m,n\leq d}$ 
defined as $L_{mn}^{(ij)}=\delta_{mj}\delta_{ni}/\sqrt d$. 
In $d=3$, 
we introduce three kinds of sets of measurement operators 
constructed from $(L_{ij})_{i,j=1}^{3}$ 
and the index sets shown in Table \ref{CompD3}. 
\begin{eqnarray*}
M_1^{(1)}=\frac{1}{\sqrt 3}
\left( \hspace{-\arraycolsep}
\begin{array}{ccc}
1 & 0 & 0\\
1 & 0 & 0\\
1 & 0 & 0
\end{array}
\right),
M_2^{(1)}=\frac{1}{\sqrt 3}
\left( \hspace{-\arraycolsep}
\begin{array}{ccc}
0 & 1 & 0\\
0 & 1 & 0\\
0 & 1 & 0
\end{array}
\right),\\
M_3^{(1)}=\frac{1}{\sqrt 3}
\left( \hspace{-\arraycolsep}
\begin{array}{ccc}
0 & 0 & 1\\
0 & 0 & 1\\
0 & 0 & 1
\end{array}
\right),\\
M_1^{(2)}=\frac{1}{\sqrt 3}
\left( \hspace{-\arraycolsep}
\begin{array}{ccc}
1 & 0 & 0\\
0 & 1 & 0\\
0 & 0 & 1
\end{array}
\right),
M_2^{(2)}=\frac{1}{\sqrt 3}
\left( \hspace{-\arraycolsep}
\begin{array}{ccc}
0 & 0 & 1\\
1 & 0 & 0\\
0 & 1 & 0
\end{array}
\right),\\
M_3^{(2)}=\frac{1}{\sqrt 3}
\left( \hspace{-\arraycolsep}
\begin{array}{ccc}
0 & 1 & 0\\
0 & 0 & 1\\
1 & 0 & 0
\end{array}
\right),
\\
M_1^{(3)}=\frac{1}{\sqrt 3}
\left( \hspace{-\arraycolsep}
\begin{array}{ccc}
1 & 0 & 0\\
0 & 0 & 1\\
0 & 1 & 0
\end{array}
\right),
M_2^{(3)}=\frac{1}{\sqrt 3}
\left( \hspace{-\arraycolsep}
\begin{array}{ccc}
0 & 1 & 0\\
1 & 0 & 0\\
0 & 0 & 1
\end{array}
\right),\\
M_3^{(3)}=\frac{1}{\sqrt 3}
\left( \hspace{-\arraycolsep}
\begin{array}{ccc}
0 & 0 & 1\\
0 & 1 & 0\\
1 & 0 & 0
\end{array}
\right).
\end{eqnarray*}

\begin{table}[tttt]
\caption{The relationship between the measurement operators and the index sets}\label{CompD3}
\begin{center}
\begin{tabular}{|c|c|c|}
\hline
$J$ & $i$ & $X^{(J,i)}$ \\ \hline
1 & 1 & $\{(1,1),(1,2),(1,3)\}$ \\
1 & 2 & $\{(2,1),(2,2),(2,3)\}$ \\
1 & 3 & $\{(3,1),(3,2),(3,3)\}$ \\
2 & 1 & $\{(1,1),(2,2),(3,3)\}$ \\
2 & 2 & $\{(1,2),(2,3),(3,1)\}$ \\
2 & 3 & $\{(1,3),(2,1),(3,2)\}$ \\
3 & 1 & $\{(1,1),(2,3),(3,2)\}$ \\
3 & 2 & $\{(1,2),(2,1),(3,3)\}$ \\
3 & 3 & $\{(1,3),(2,2),(3,1)\}$ \\
\hline
\end{tabular}
\end{center}
\end{table}

\end{document}